\documentclass[pre]{revtex4}% Physical Review B

\usepackage{graphicx}% Include figure files
\usepackage{dcolumn}% Align table columns on decimal point
\usepackage{bm}% bold math
\usepackage{subfigure}
\usepackage{amssymb}
\usepackage{graphics}
\usepackage{epsfig}
\usepackage{threeparttable}
\newcommand{\thickhline}{\noalign{\hrule height 0.8pt}}

\usepackage{threeparttable}

\begin{document}

\preprint{APS/123-QED}

\title{Effective field theory analysis of $3D$ random field Ising model on isometric lattices}% Force line breaks with \\
\author{\"{U}mit Ak{\i}nc{\i}}
\author{Yusuf Y\"{u}ksel}%
\altaffiliation[Also at ]{Dokuz Eyl\"{u}l University,
Graduate School of Natural and Applied Sciences, Turkey}%Lines break automatically or can be forced with \\
\affiliation{Department of Physics, Dokuz Eyl\"{u}l University,
TR-35160 Izmir, Turkey}
\author{Hamza Polat}
\email{hamza.polat@deu.edu.tr}
\affiliation{Department of Physics, Dokuz Eyl\"{u}l University,
TR-35160 Izmir, Turkey}
%\affiliation{%


\date{\today}% It is always \today, today,

\begin{abstract}
Ising model with quenched random magnetic fields is examined for single Gaussian, bimodal and double Gaussian random field distributions by introducing an effective field approximation that takes into account the correlations between different spins that emerge when expanding the identities. Random field distribution shape dependencies of the phase diagrams and magnetization curves are investigated for simple cubic, body centered and face centered cubic lattices. The conditions for the occurrence of reentrant behavior and tricritical points on the system are also discussed in detail.
\begin{description}
\item[\qquad \qquad \qquad PACS numbers]
75.10.Hk, 75.30.Kz, 75.50.Lk
\end{description}
\end{abstract}

\maketitle
\tableofcontents

\section{Introduction}\label{introduction}
Ising model \cite{ising,onsager} which was originally introduced as a model describing the phase transition properties of ferromagnetic materials has been widely examined in statistical mechanics and condensed matter physics. In the course of time, basic conclusions of this simple model has been improved by introducing new concepts such as disorder effects on the critical behavior of the systems in question. Ising model in a quenched random field (RFIM) which has been studied over three decades is an example of this situation. The model which is actually based on the local fields acting on the lattice sites which are taken to be random according to a given probability distribution was introduced for the first time by Larkin \cite{larkin} for superconductors and later generalized by Imry and Ma \cite{imry_ma}. Lower critical dimension $d_{c}$ of RFIM has been remained as an unsolved mystery for many years and various theoretical methods have been introduced. For example, domain wall argument of Imry and Ma \cite{imry_ma} suggests that a transition should exist in three and higher dimensions for finite temperature and randomness which means that $d_{c}=2$ \cite{grinstein,fernandez,imbrie,bricmont}. On the contrary, dimensional reduction arguments \cite{parisi} conclude that the system should not have a phase transition at finite temperature in three dimensions or fewer, so $d_{c}=3$ \cite{binder,mukamel2,mukamel,niemi}. On the other hand, Frontera and Vives \cite{frontera} showed that a two dimensional ferromagnetic RFIM with Gaussian random field distribution exhibits order at zero temperature.

At the same time, a great many of theoretical and experimental works have paid attention on the RFIM and quite noteworthy results have been obtained. For instance, it has been shown that diluted antiferromagnets such as $\mathrm{Fe}_{x}\mathrm{Zn}_{1-x}\mathrm{F}_{2}$\cite{belanger,king}, $\mathrm{Rb}_{2}\mathrm{Co}_{x}\mathrm{Mg}_{1-x}\mathrm{F}_{4}$\cite{ferreira,yoshizawa} and $\mathrm{Co}_{x}\mathrm{Zn}_{1-x}\mathrm{F}_{2}$\cite{yoshizawa}  in a uniform magnetic field just correspond to a ferromagnet in a random uniaxial magnetic field \cite{fishman,cardy}. Following studies have been devoted to investigate the phase diagrams of these systems in depth and in the mean field level it was found that different random field distributions lead to different phase diagrams for infinite dimensional models. For example, using a Gaussian probability distribution Schneider and Pytte\cite{schneider_pytte} have shown that phase diagrams of the model exhibit only second order phase transition properties. On the other hand, Aharony \cite{aharony} and Matthis \cite{matthis} have introduced bimodal and trimodal distributions, respectively and they have reported the observation of tricritical behavior. In a recent series of papers, phase transition properties of infinite dimensional RFIM with symmetric double \cite{crokidakis} and triple \cite{salmon} Gaussian random fields have also been studied by means of a replica method and a rich variety of phase diagrams have been presented. The situation has also been handled on three dimensional lattices with nearest neighbor interactions by a variety of theoretical works such as effective field theory (EFT) \cite{borges,liang2,sarmento,sebastianes,kaneyoshi}, Monte Carlo (MC) simulations \cite{landau,machta,fytas1,fytas2}, pair approximation (PA) \cite{albayrak} and series expansion (SE) method \cite{gofman}. By using EFT, Borges and Silva \cite{borges} studied the system for square $(q=4)$ and simple cubic $(q=6)$ lattices and they observed a tricritical point only for $(q\geq6)$. Similarly, Sarmento and Kaneyoshi \cite{sarmento} investigated the phase diagrams of RFIM by means of EFT with correlations for a bimodal field distribution and they concluded that reentrant behavior of second order is possible for the system with $(q\geq6)$. Recently, Fytas et al. \cite{fytas1} applied Monte Carlo simulations on a simple cubic lattice. They found that the transition remains continuous for a bimodal field distribution, while Hadjiagapiou \cite{hadjiagapiou} observed reentrant behavior and confirmed the existence of tricritical point for an asymmetric bimodal probability distribution within the mean-field approximation based on Landau expansion.

Conventional EFT approximations include spin-spin correlations resulting from the usage of the Van der Waerden identities and provide results that are superior to those obtained within the traditional MFT. However, these conventional EFT approximations are not sufficient enough to improve the results, due to the usage of a decoupling approximation (DA) that neglects the correlations between different spins that emerge when expanding the identities. Therefore, taking these correlations into consideration will improve the results of conventional EFT approximations. In order to overcome this point, recently we proposed an approximation that takes into account the correlations between different spins in the cluster of considered lattice \cite{polat,canpolat,yuksel_1,yuksel_2,yuksel_3}. Namely, an advantage of the approximation method proposed by these studies is that no decoupling procedure is used for the higher order correlation functions. On the other hand, as far as we know, EFT studies in the literature dealing with RFIM are based only on  discrete probability distributions (bimodal or trimodal). Hence, in this work we intent to study the phase diagrams of the RFIM with single Gaussian, bimodal and double Gaussian random field distributions on isometric lattices.

Organization of the paper is as follows: In section \ref{formulation} we briefly present the formulations. The results and discussions are presented in section \ref{results}, and finally section \ref{conclude} contains our conclusions.

\section{Formulation}\label{formulation}
In this section, we give the formulation of present study for simple cubic lattice with $q=6$. A detailed explanation of the method for body centered and face centered cubic lattices can be found in Appendix. As a simple cubic lattice, we consider a three-dimensional lattice which has $N$ identical spins arranged. We define a
cluster on the lattice which consists of a central spin labeled $S_{0}$, and $q$ perimeter spins being the
nearest neighbors of the central spin. The cluster consists of $(q+1)$ spins being independent
from the spin operator $\hat{S}$. The nearest neighbor spins are in an effective field produced by the outer
spins, which can be determined by the condition that the thermal average of the central spin is
equal to that of its nearest neighbor spins. The Hamiltonian describing our model is
\begin{equation}\label{eq1}
H=-J\sum_{<i,j>}S_{i}^{z}S_{j}^{z}-\sum_{i}h_{i}S_{i}^{z},
\end{equation}
where the first term is a summation over the nearest neighbor spins with $S_{i}^{z}=\pm1$ and
the second term represents the Zeeman interactions on the lattice. Random magnetic fields are distributed according to a given probability distribution function.
Present study deals with three kinds of field distribution. Namely, a normal distribution which is defined as
\begin{equation}\label{eq2b}
P(h_{i})=\left(\frac{1}{2\pi\sigma^{2}}\right)^{1/2}\exp\left[-\frac{h_{i}^{2}}{2\sigma^{2}}\right],
\end{equation}
with a width $\sigma$ and zero mean, a bimodal discrete distribution
\begin{equation}\label{eq2a}
P(h_{i})=\frac{1}{2}\left[\delta(h_{i}-h_{0})+\delta(h_{i}+h_{0})\right],
\end{equation}
where half of the lattice sites subject to a magnetic field $h_{0}$ and the remaining lattice sites have a field $-h_{0}$,  and a double peaked Gaussian distribution
\begin{equation}\label{eq2c}
P(h_{i})=\frac{1}{2}\left(\frac{1}{2\pi\sigma^{2}}\right)^{1/2}\left\{\exp\left[-\frac{(h_{i}-h_{0})^{2}}{2\sigma^{2}}\right]+ \exp\left[-\frac{(h_{i}+h_{0})^{2}}{2\sigma^{2}}\right]\right\}.
\end{equation}
In a double peaked distribution defined in equation (\ref{eq2c}), random fields $\pm h_{0}$ are distributed with equal probability and the form of the distribution depends on the $h_{0}$ and $\sigma$ parameters, where $\sigma$ is the width of the distribution.

According to Callen identity \cite{callen} for the spin-1/2 Ising ferromagnetic system, the thermal average of
the spin variables at the site $i$ is given by
\begin{equation}\label{eq3}
\left\langle \{f_{i}\}S_{i}^{z}\right\rangle=\left\langle \{f_{i}\}\tanh\left[\beta\left(J\sum_{j} S_{j}+h_{i}\right)\right]\right\rangle,
\end{equation}
where $j$ expresses the nearest neighbor sites of the central spin and $\{f_{i}\}$ can be any function of the Ising variables as long as it is not a function of the site. From equation (\ref{eq3}) with $f_{i}=1$, the thermal and random-configurational averages of a central spin can be represented on a simple cubic lattice by introducing the differential operator technique \cite{honmura_kaneyoshi,kaneyoshi_1}
\begin{equation}\label{eq4}
m_{0}=\left\langle\left\langle S_{0}^{z}\right\rangle\right\rangle_{r}=\left\langle\left\langle \prod_{j=1}^{q=6}\left[\cosh(J\nabla)+S_{j}^{z}\sinh(J\nabla)\right]\right\rangle\right\rangle_{r} F(x)|_{x=0},
\end{equation}
where $\nabla$ is a differential operator, $q$ is the coordination number of the lattice and the inner $\langle...\rangle$ and the outer $\langle...\rangle_{r}$ brackets represent the thermal and configurational averages, respectively. The function $F(x)$ in equation (\ref{eq4}) is defined by
\begin{equation}\label{eq5}
F(x)=\int dh_{i}P(h_{i})\tanh[\beta(x+h_{i})],
\end{equation}
and it has been calculated by numerical integration and by using the distribution functions defined in equations (\ref{eq2b}), (\ref{eq2a}) and (\ref{eq2c}).
By expanding the right hand side of equation (\ref{eq4}) we get the longitudinal spin correlation as
\begin{eqnarray}\label{eq6}
\nonumber\langle\langle S_{0}\rangle\rangle_{r}&=&k_{0}+6k_{1}\langle\langle S_{1}\rangle\rangle_{r}+15k_{2}\langle\langle S_{1}S_{2}\rangle\rangle_{r}+20k_{3}\langle\langle S_{1}S_{2}S_{3}\rangle\rangle_{r}+15k_{4}\langle\langle S_{1}S_{2}S_{3}S_{4}\rangle\rangle_{r}\\
&&+6k_{5}\langle\langle S_{1}S_{2}S_{3}S_{4}S_{5}\rangle\rangle_{r}+k_{6}\langle\langle S_{1}S_{2}S_{3}S_{4}S_{5}S_{6} \rangle\rangle_{r}.
\end{eqnarray}
The coefficients in equation (\ref{eq6}) are defined as follows
\begin{eqnarray}\label{eq7}
\nonumber
k_{0}&=&\cosh^{6}(J\nabla)F(x)|_{x=0},\\
\nonumber
k_{1}&=&\cosh^{5}(J\nabla)\sinh(J\nabla)F(x)|_{x=0},\\
\nonumber
k_{2}&=&\cosh^{4}(J\nabla)\sinh^{2}(J\nabla)F(x)|_{x=0},\\
\nonumber
k_{3}&=&\cosh^{3}(J\nabla)\sinh^{3}(J\nabla)F(x)|_{x=0},\\
\nonumber
k_{4}&=&\cosh^{2}(J\nabla)\sinh^{4}(J\nabla)F(x)|_{x=0},\\
\nonumber
k_{5}&=&\cosh(J\nabla)\sinh^{5}(J\nabla)F(x)|_{x=0},\\
k_{6}&=&\sinh^{6}(J\nabla)F(x)|_{x=0}.
\end{eqnarray}
Next, the average value of the perimeter spin in the system can be written as follows and it is found as
\begin{eqnarray}\label{eq8}
\nonumber
m_{1}=\langle\langle S_{1}\rangle\rangle_{r} &=&\langle\langle \cosh(J\nabla)+S_{0}\sinh(J\nabla)\rangle\rangle_{r} F(x+\gamma),\\
&=&a_{1}+a_{2}\langle\langle S_{0}\rangle\rangle_{r}.
\end{eqnarray}
For the sake of simplicity, the superscript $z$ is omitted from the left and right hand sides of equations (\ref{eq6}) and (\ref{eq8}).
The coefficients in equation (\ref{eq8}) are defined as
\begin{eqnarray}\label{eq9}
\nonumber
a_{1}&=&\cosh(J\nabla)F(x+\gamma)|_{x=0},\\
a_{2}&=&\sinh(J\nabla)F(x+\gamma)|_{x=0}.
\end{eqnarray}
In equation (\ref{eq9}), $\gamma=(q -1)A$ is the effective field produced by the $(q-1)$ spins outside of the system and $A$ is an unknown parameter to be determined self-consistently. Equations (\ref{eq6}) and (\ref{eq8}) are the fundamental correlation functions of the system. When the right-hand side of equation (\ref{eq4}) is expanded, the multispin correlation functions appear. The simplest approximation, and one of the most frequently adopted is to decouple these correlations according to
\begin{equation}\label{eq10}
\left\langle\left\langle
S_{i}S_{j}...S_l\right\rangle\right\rangle_{r}\cong\left\langle\left\langle
S_i\right\rangle\right\rangle_{r}\left\langle\left\langle
S_j\right\rangle\right\rangle_{r}...\left\langle\left\langle
S_l\right\rangle\right\rangle_{r},
\end{equation}
for $i\neq j \neq...\neq l$ \cite{tamura_kaneyoshi}. The main difference of the method used in this study from the other approximations in the literature emerges in comparison with any decoupling approximation (DA) when expanding the right-hand side of
equation (\ref{eq4}). In other words, one advantage of the approximation method used in this study is that such a kind of decoupling procedure is not used for the higher order correlation functions.
For spin-1/2 Ising system in a random field, taking equations (\ref{eq6}) and (\ref{eq8}) as a basis we derive a set of linear equations of the spin correlation functions in the system. At this point, we assume that $(i)$ the correlations depend only on the distance between the spins and $(ii)$ the average values of a central spin and its nearest-neighbor spin (it is labeled as the perimeter spin) are equal to each other with the fact that, in the matrix representations of spin operator $\hat{S}$, the spin-1/2 system has the property $(\hat{S})^{2}=1$ . Thus, the number of linear equations obtained for a simple cubic lattice $(q=6)$ reduces to twelve and the complete set is as follows
\begin{eqnarray}\label{eq11}
\nonumber
\left\langle\left\langle S_{1}\right\rangle\right\rangle_{r}&=&a_{1}+a_{2}\langle\langle S_{0}\rangle\rangle_{r},\\
\nonumber
\left\langle\left\langle S_{1}S_{2}\right\rangle\right\rangle_{r}&=&a_{1}\langle\langle S_{1}\rangle\rangle_{r}+a_{2}\langle\langle S_{0}S_{1}\rangle\rangle_{r},\\
\nonumber
\left\langle\left\langle S_{1}S_{2}S_{3}\right\rangle\right\rangle_{r}&=&a_{1}\langle\langle S_{1}S_{2}\rangle\rangle_{r}+a_{2}\langle\langle S_{0}S_{1}S_{2}\rangle\rangle_{r},\\
\nonumber
\left\langle\left\langle S_{1}S_{2}S_{3}S_{4}\right\rangle\right\rangle_{r}&=&a_{1}\langle\langle S_{1}S_{2}S_{3}\rangle\rangle_{r}+a_{2}\langle\langle S_{0}S_{1}S_{2}S_{3}\rangle\rangle_{r},\\
\nonumber
\left\langle\left\langle S_{1}S_{2}S_{3}S_{4}S_{5}\right\rangle\right\rangle_{r}&=&a_{1}\langle\langle S_{1}S_{2}S_{3}S_{4}\rangle\rangle_{r}+a_{2}\langle\langle S_{0}S_{1}S_{2}S_{3}S_{4}\rangle\rangle_{r},\\
\nonumber
\left\langle\left\langle S_{1}S_{2}S_{3}S_{4}S_{5}S_{6}\right\rangle\right\rangle_{r}&=&a_{1}\langle\langle S_{1}S_{2}S_{3}S_{4}S_{5}\rangle\rangle_{r}+a_{2}\langle\langle S_{0}S_{1}S_{2}S_{3}S_{4}S_{5}\rangle\rangle_{r},\\
\nonumber\langle\langle S_{0}\rangle\rangle_{r}&=&k_{0}+6k_{1}\langle\langle S_{1}\rangle\rangle_{r}+15k_{2}\langle\langle S_{1}S_{2}\rangle\rangle_{r}+20k_{3}\langle\langle S_{1}S_{2}S_{3}\rangle\rangle_{r}+15k_{4}\langle\langle S_{1}S_{2}S_{3}S_{4}\rangle\rangle_{r}\\
\nonumber&&+6k_{5}\langle\langle S_{1}S_{2}S_{3}S_{4}S_{5}\rangle\rangle_{r}+k_{6}\langle\langle S_{1}S_{2}S_{3}S_{4}S_{5}S_{6} \rangle\rangle_{r},\\
\nonumber
\left\langle\left\langle S_{0}S_{1}\right\rangle\right\rangle_{r}&=&6k_{1}+(k_{0}+15k_{2})\langle\langle S_{1}\rangle\rangle_{r}+20k_{3}\langle\langle S_{1}S_{2}\rangle\rangle_{r}+15k_{4}\langle\langle S_{1}S_{2}S_{3}\rangle\rangle_{r}\\
\nonumber&&+6k_{5}\langle\langle S_{1}S_{2}S_{3}S_{4}\rangle\rangle_{r}+k_{6}\langle\langle S_{1}S_{2}S_{3}S_{4}S_{5} \rangle\rangle_{r},\\
\nonumber
\left\langle\left\langle S_{0}S_{1}S_{2}\right\rangle\right\rangle_{r}&=&(6k_{1}+20k_{3})\langle\langle S_{1}\rangle\rangle_{r}+(k_{0}+15k_{2}+15k_{4})\langle\langle S_{1}S_{2} \rangle\rangle_{r}\\
\nonumber&&+6k_{5}\langle\langle S_{1}S_{2}S_{3}\rangle\rangle_{r}+k_{6}\langle\langle S_{1}S_{2}S_{3}S_{4}\rangle\rangle_{r},\\
\nonumber
\left\langle\left\langle S_{0}S_{1}S_{2}S_{3}\right\rangle\right\rangle_{r}&=&(6k_{1}+20k_{3}+6k_{5})\langle\langle S_{1}S_{2}\rangle\rangle_{r}+(k_{0}+15k_{2}+15k_{4}+k_{6})\langle\langle S_{1}S_{2}S_{3}\rangle\rangle_{r},\\
\nonumber
\left\langle\left\langle S_{0}S_{1}S_{2}S_{3}S_{4}\right\rangle\right\rangle_{r}&=&(6k_{1}+20k_{3}+6k_{5})\langle\langle S_{1}S_{2}S_{3}\rangle\rangle_{r}+(k_{0}+15k_{2}+15k_{4}+k_{6})\langle\langle S_{1}S_{2}S_{3}S_{4}\rangle\rangle_{r},\\
\left\langle\left\langle S_{0}S_{1}S_{2}S_{3}S_{4}S_{5}\right\rangle\right\rangle_{r}&=&(6k_{1}+20k_{3}+6k_{5})\langle\langle S_{1}S_{2}S_{3}S_{4}\rangle\rangle_{r}+(k_{0}+15k_{2}+15k_{4}+k_{6})\langle\langle S_{1}S_{2}S_{3}S_{4}S_{5}\rangle\rangle_{r}
\end{eqnarray}
If equation (\ref{eq11}) is written in the form of a $12\times12$ matrix and solved in terms of the variables $x_{i}[(i=1,2,...,12)(e.g., x_{1}=\langle\langle S_{1}\rangle\rangle_{r}, x_{2}=\langle\langle S_{1}S_{2}\rangle\rangle_{r},x_{3}=\langle\langle S_{1}S_{2}S_{3}\rangle\rangle_{r},...)]$ of the linear equations, all of the spin correlation functions can be easily determined as functions of the temperature and Hamiltonian/random field parameters. Since the thermal and configurational average of the central spin is equal to that of its nearest-neighbor spins within the present method, the unknown parameter $A$ can be numerically determined by the relation
\begin{equation}\label{eq12}
\langle\langle S_{0}\rangle\rangle_{r}=\langle\langle
S_{1}\rangle\rangle_{r} \qquad {\rm{ or }}\qquad x_{7}=x_{1}.
\end{equation}
By solving equation (\ref{eq12}) numerically at a given fixed set of Hamiltonian/random field parameters we obtain the parameter $A$. Then we use the numerical values of $A$ to obtain the spin correlation functions which can be found from equation (\ref{eq11}). Note that $A=0$ is always the root of equation (\ref{eq12}) corresponding to the disordered state of the system. The nonzero root of $A$ in equation (\ref{eq12}) corresponds to the long range ordered state of the system. Once the spin correlation functions have been evaluated then we can give the numerical results for the thermal and magnetic properties of the system. Since the effective field $\gamma$ is very small in the vicinity of $k_{B}T_{c}/J$, we can obtain the critical temperature for the fixed set of Hamiltonian/random field parameters by solving equation (\ref{eq12}) in the limit of $\gamma\rightarrow0$ then we can construct the whole phase diagrams of the system. Depending on the Hamiltonian/random field parameters, there may be two solutions (i.e. two critical temperature values satisfy the equation (\ref{eq12})) corresponding to the first/second and second order phase transition points, respectively. We determine the type of the transition by looking at the temperature dependence of magnetization for selected values of system parameters.

\section{Results and Discussion}\label{results}

In this section, we discuss how the type of random field distribution effects the phase diagrams of the system. Also, in order to clarify the type of the transitions in the system, we give the temperature dependence of the order parameter.

\subsection{Phase diagrams of single Gaussian distribution}
The form of single Gaussian distribution which is defined in equation (\ref{eq2b}) is governed by only one parameter $\sigma$ which is the width of the distribution. In Fig. 1, we show the phase diagram of the system for simple cubic, body centered cubic and face centered cubic lattices in $(k_{B}T_{c}/J-\sigma)$ plane. We can clearly see that as $\sigma$ increases then the width of the distribution function gets wider and randomness effect of magnetic field distribution on the system becomes significantly important. Therefore, increasing $\sigma$ value causes a decline in the critical temperature $k_{B}T_{c}/J$ of the system. We note that the critical temperature of the system reaches to zero at $\sigma=3.8501$, $5.450$ and $8.601$ for simple cubic $(q=6)$, body centered $(q=8)$ and face centered cubic $(q=12)$ lattices, respectively. Besides, we have not observed any reentrant/tricritical behavior for single Gaussian distribution, or in other words, the system undergoes only a second order phase transition. $k_{B}T_{c}/J$ value in the absence of any randomness i.e. when $\sigma=0$ is obtained as $k_{B}T_{c}/J=4.5274$, $6.5157$ and $10.4986$ for $q=6, 8, 12$, respectively. These values can be compared with the other works in the literature. In Table 1, we see that present work improves the results of the other works. Hence, it is obvious that the present method is superior to the other EFT methods in the literature. The reason is due to the fact that, in contrast to the previously published works mentioned above, there is no uncontrolled decoupling procedure used for the higher order correlation functions within the present approximation.
\begin{table}[h]\label{table1}
%\begin{flushleft}
\begin{center}
\begin{threeparttable}
\caption{Critical temperature $k_{B}T_{c}/J$ at $h_{0}/J=0$ and $\sigma=0$ obtained by several methods and the present work for $q=6,8,12$.}
\renewcommand{\arraystretch}{1.3}
\begin{tabular}{llllllllllll}
\thickhline
Lattice & EBPA\cite{du} & CEFT\cite{kaneyoshi_2} & \ \ PA\cite{balcerzak} & \ EFT\cite{kaneyoshi_3} & \ \ BA\cite{kikuchi}&EFRG\cite{sousa}&MFRG\cite{reinerhr} & MC\cite{landau2} & \ \ SE\cite{fisher} & Present Work\\
\hline SC  & 4.8108 & \ 4.9326 & \ 4.9328 & \ \ 5.0732  & \ \ 4.6097 & \ 4.85& \ \ 4.93 & \ 4.51 & \ 4.5103 & \ \ \ 4.5274  \\
BCC   &\ \ \ - &\ 6.9521 & \ \ \ \ - & \ \ \ - \  &\ \ \ \ - \ &\ 6.88 \ & \ \ 6.95 & \ 6.36 & \ 6.3508 & \ \ \ 6.5157 \\
FCC   & \ \ \ - &\ 10.9696 & \ \ \ \ - & \ \ \ - \  &\ \ \ \ - \ &\ \ \ - \ & \ \ \ \ - & \ \ - & \ 9.7944 & \ \ \ 10.4986 \\
\thickhline \\
\end{tabular}
%\end{flushleft}
\end{threeparttable}
\end{center}
\end{table}

Magnetization surfaces and their projections on $(k_{B}T/J-\sigma)$ plane corresponding to the phase diagrams shown in Fig. 1 are depicted in Fig. 2 with $q=6, 8$ and $12$. We see that as the temperature increases starting from zero, magnetization of the system decreases continuously and it falls rapidly to zero at the critical temperature for all $\sigma$ values. Moreover, critical temperature of the system decreases and saturation value of magnetization curves remains constant for a while then reduces as $\sigma$ value increases. This is a reasonable result since, if $\sigma$ value increases, then the randomness effects increasingly play an important role on the system and random fields have a tendency to destruct the long range ferromagnetic order on the system, and hence magnetization weakens. These observations are common properties of both three lattices.

\begin{figure}[!h]
\includegraphics[width=8cm]{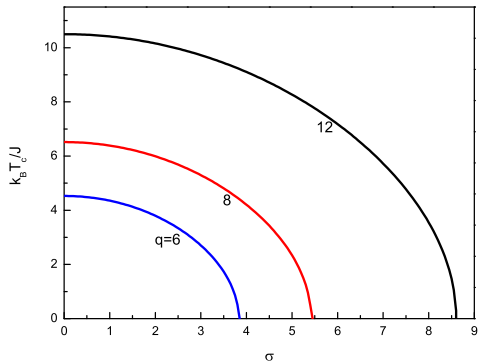}\\
\caption{Phase diagrams of simple cubic $(q=6)$, body centered cubic $(q=8)$ and face centered cubic $(q=12)$ lattices in $(k_{B}T_{c}/J-\sigma)$ plane corresponding to single Gaussian distribution of random fields. The numbers on each curve denote the coordination numbers.}\label{fig1}
\end{figure}

\begin{figure}[!h]
\includegraphics[width=15cm]{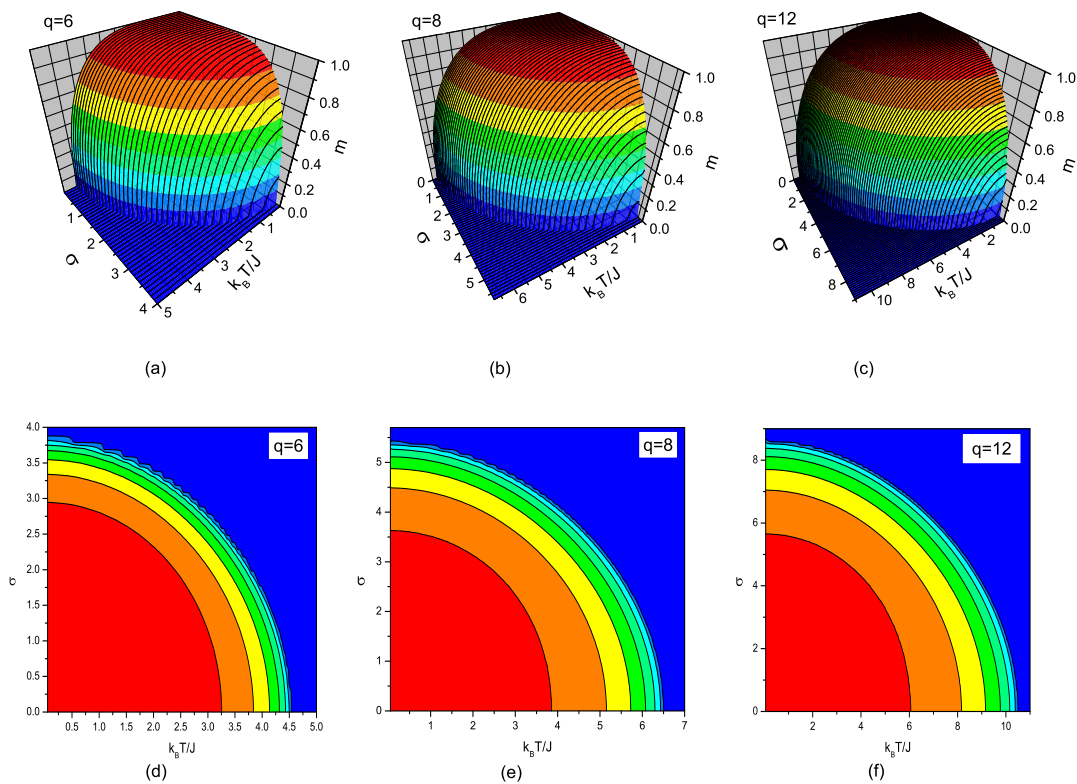}\\
\caption{Variation of magnetization with $k_{B}T/J$ and $\sigma$ corresponding to the phase diagrams in Fig. 1 for simple cubic, body centered and face centered cubic lattices. (a)-(c) 3D contour plot surfaces, (d)-(f) projections on $(k_{B}T/J-\sigma)$ plane.}\label{fig2}
\end{figure}

\subsection{Phase diagrams of bimodal distribution}
Next, in order to investigate the effect of the bimodal random fields defined in equation (\ref{eq2a}) on the phase diagrams of the system, we show the phase diagrams in $(k_{B}T_{c}/J-h_{0}/J)$ plane and corresponding magnetization profiles with coordination numbers $q=6, 8, 12$ in Figs. 3 and 4. In these figures solid and dashed lines correspond to second and first order phase transition points, respectively and hollow circles in Fig. 3 denote tricritical points. As seen in Fig. 3, increasing values of $h_{0}/J$ causes the critical temperature to decrease for a while, then reentrant behavior of first order occurs at a specific range of $h_{0}/J$. According to our calculations, reentrant phenomena and the first order phase transitions can be observed within the range of $2.0<h_{0}/J<3.0$ for $q=6$, $3.565<h_{0}/J<3.95$ for $q=8$ and $4.622<h_{0}/J<5.81$ for $q=12$. If $h_{0}/J$ value is greater than the upper limits of these  field ranges, the system exhibits no phase transition. The tricritical temperatures $k_{B}T_{t}/J$ which are shown as hollow circles in Fig. 2 are found as $k_{B}T_{t}/J=1.5687, 2.4751$ and $4.3769$ for $q=6,8,12$, respectively.
\begin{figure}[!h]
\includegraphics[width=8cm]{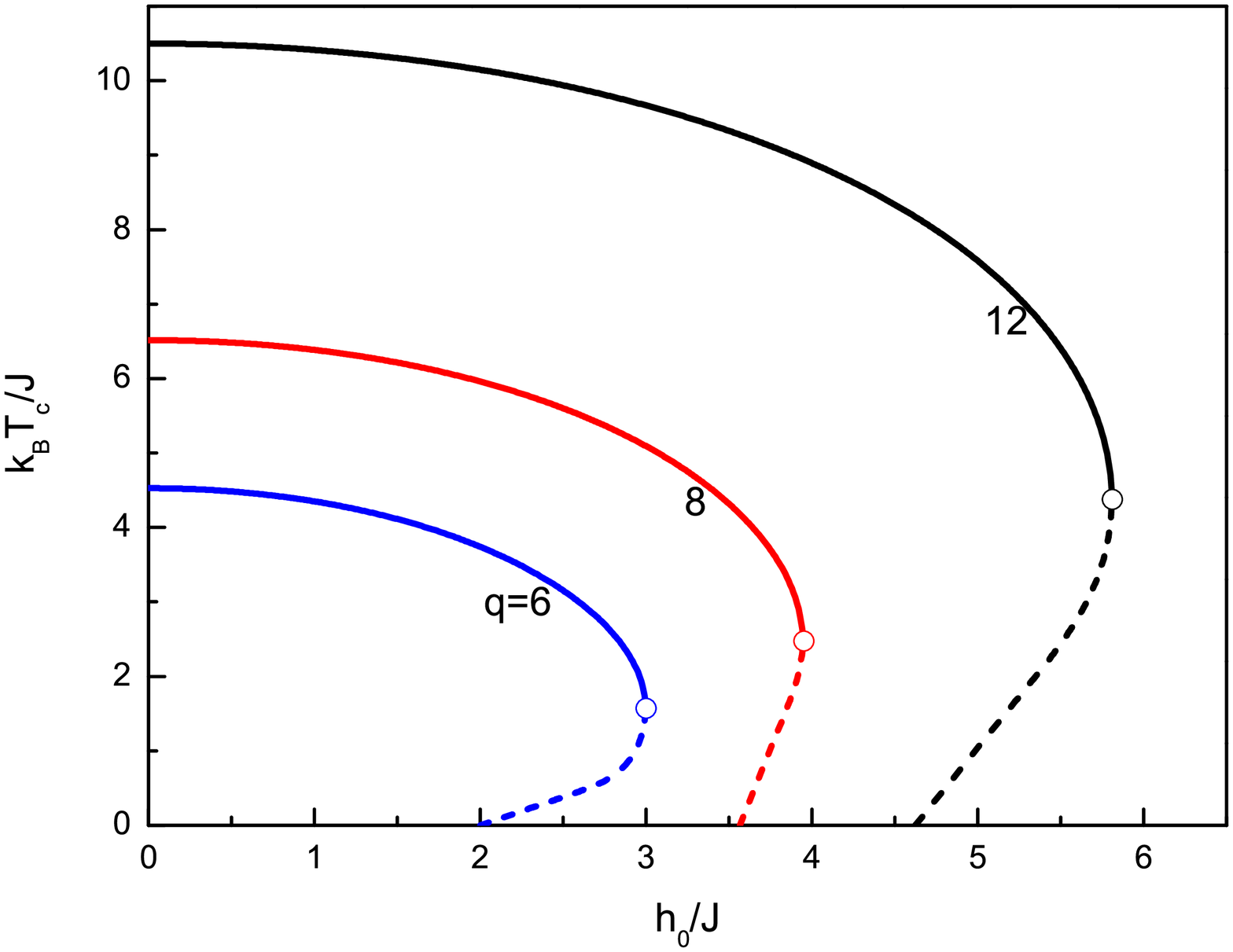}\\
\caption{Phase diagrams of the system $(q=6,8,12)$ in $(k_{B}T_{c}/J-h_{0}/J)$ plane, corresponding to bimodal random field distribution. Solid and dashed lines correspond to second and first order phase transitions, respectively. Hollow circles denote the tricritical points.}\label{fig3}
\end{figure}
\begin{figure}[!h]
\includegraphics[width=15cm]{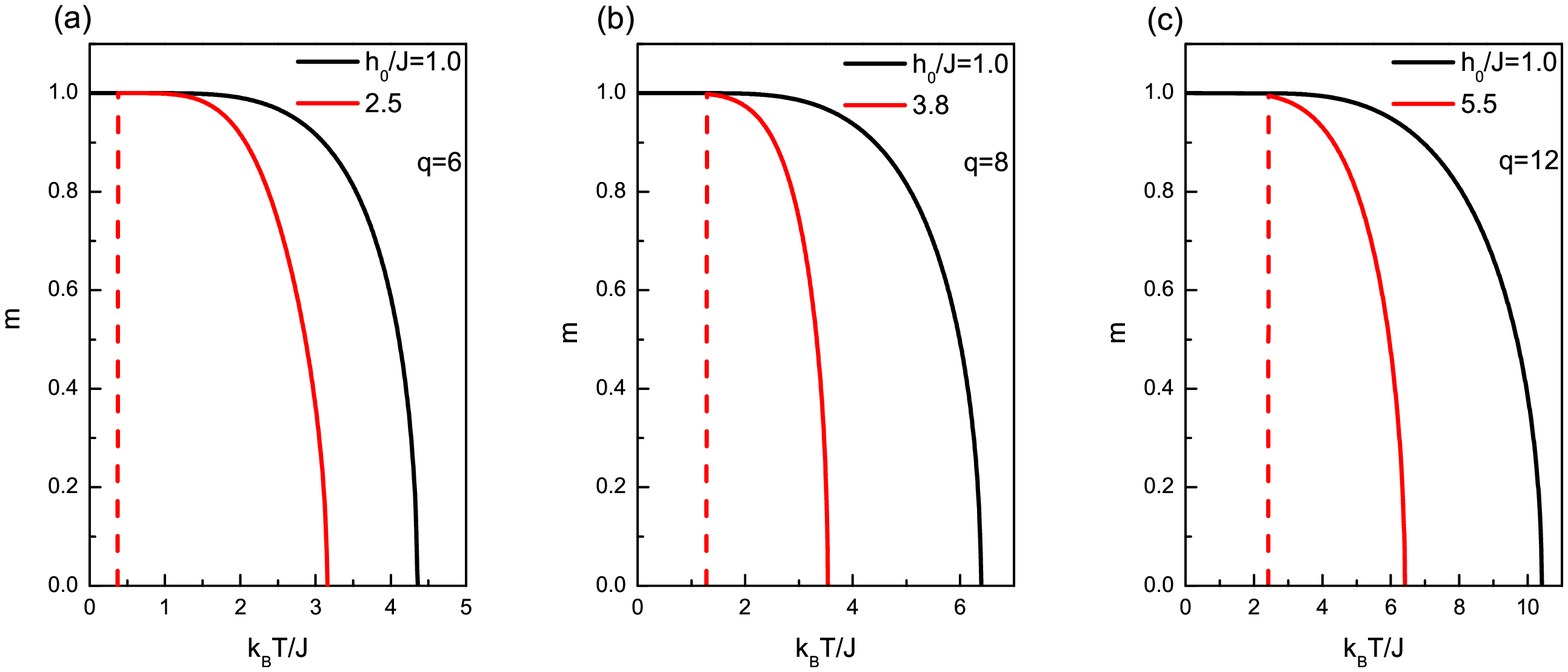}\\
\caption{Temperature dependence of magnetization corresponding to Fig. 3 with $\sigma=0$  and $h_{0}/J=1.0,2.5$ for simple cubic lattice (left panel), $h_{0}/J=1.0,3.8$ for body centered cubic lattice (middle panel) and $h_{0}/J=1.0,5.5$ for face centered cubic lattice (right panel). Solid and dashed lines correspond to second and first order phase transitions, respectively.}\label{fig4}
\end{figure}

In Fig. 4, we show two typical magnetization profiles of the system. Namely, the system always undergoes a second order phase transition for $h_{0}/J=1.0$. On the other hand, two successive phase transitions (i.e. a first order transition is followed by a second order phase transition) occur at the values of $h_{0}/J=2.5, 3.8$ and $5.5$  for $q=6, 8$ and $12$, respectively, which puts forward the existence of a first order reentrant phenomena on the system. We observe that the increasing $h_{0}/J$ values do not effect the saturation values of magnetization curves.
\subsection{Phase diagrams of double Gaussian distribution}
\begin{figure}[h!]
\includegraphics[width=14cm]{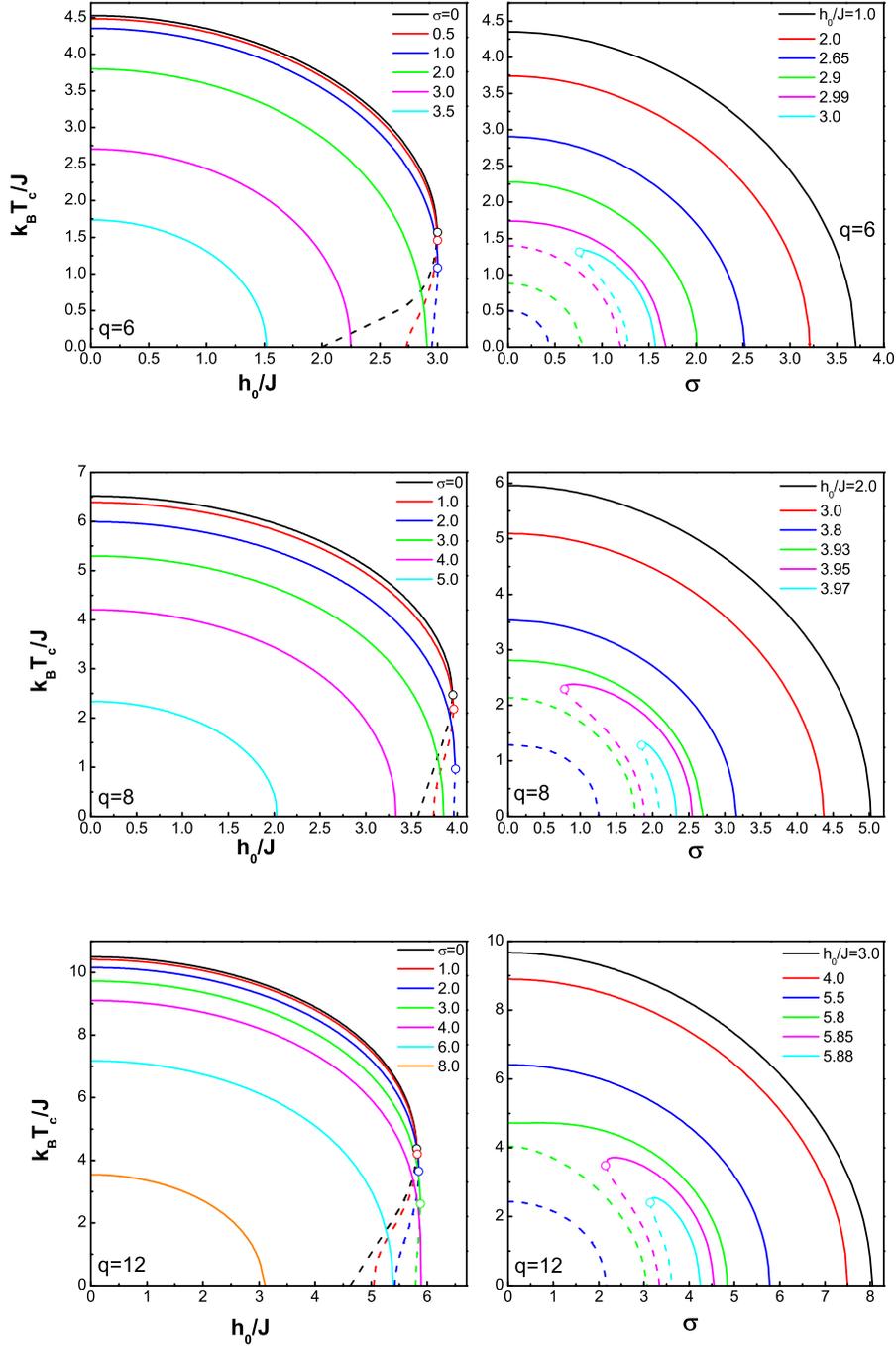}\\
\caption{Phase diagrams of the system for a double Gaussian random field distribution with $q=6,8,12$ in $(k_{B}T_{c}/J-h_{0}/J)$ and $(k_{B}T_{c}/J-\sigma)$ planes.}\label{fig5}
\end{figure}

Double Gaussian distribution in equation (\ref{eq2c}) with nearest neighbor interactions have not yet been examined in the literature. Therefore, it would be interesting to investigate the phase diagrams of the system with random fields corresponding to equation (\ref{eq2c}). Now the shape of the random fields is governed by two parameters $h_{0}/J$ and $\sigma$. As shown in preceding results, increasing $\sigma$ value tends to reduce the saturation value of the order parameter and destructs the second order phase transitions by decreasing the critical temperature of the system without exposing any reentrant phenomena for $h_{0}/J=0$. Besides, as $h_{0}/J$ value increases then the second order phase transition temperature decreases again and the system may exhibit a reentrant behavior for $\sigma=0$, while saturation value of magnetization curves remains unchanged. Hence, the presence of both $h_{0}/J$ and $\sigma$ on the system should produce a cooperative effect on the phase diagrams of the system. Fig. 5 shows the phase diagrams of the system in $(k_{B}T_{c}/J-h_{0}/J)$ and $(k_{B}T_{c}/J-\sigma)$ planes for $q=6, 8$ and $12$. As we can see on the left panels in Fig. 5, the system exhibits tricritical points and reentrant phenomena for narrow width of the random field distribution and as the width $\sigma$  of the distribution gets wider then the reentrant phenomena and tricritical behavior disappear. In other words, both the reentrant behavior and tricritical points disappear as $\sigma$ parameter becomes significantly dominant on the system. Our results predict that tricritical points depress to zero at $\sigma=1.421, 2.238$ and $3.985$ for $q=6, 8$ and $12$, respectively. For distribution widths greater than these values, all transitions are of second order and as we increase $\sigma$ value further, then ferromagnetic region gets narrower. Similarly, on the right panels in Fig. 5, we investigate the phase diagrams of the system in $(k_{B}T_{c}/J-\sigma)$ plane with selected values of $h_{0}/J$. We note that for the values of $h_{0}/J\leq 2.0$ $(q=6)$, $h_{0}/J\leq 3.565$ $(q=8)$ and $h_{0}/J\leq 4.622$ $(q=12)$, system always undergoes a second order phase transition between paramagnetic and ferromagnetic phases at a critical temperature which decreases with increasing values of $h_{0}/J$ as in Fig. 1, where $h_{0}/J=0$. Moreover, for the values of $h_{0}/J$ greater than these threshold values the system exhibits a reentrant behavior of first order and the transition lines exhibit a bulge which gets smaller with increasing values of $h_{0}/J$ which again  means that ferromagnetic phase region gets narrower. Besides, for $h_{0}/J>2.9952$ $(q=6)$, $h_{0}/J>3.9441$ $(q=8)$ and $h_{0}/J>5.8085$ $(q=12)$ tricritical points appear on the system. In Fig. 6, we show magnetization curves corresponding to the phase diagrams shown in Fig. 5 for simple cubic lattice. Fig. 6a shows the temperature dependence of magnetization curves for $q=6$ with $h_{0}/J=0.5$ (left panel) and $h_{0}/J=2.5$ (right panel) with selected values of $\sigma$. As we can see in Fig. 6a, as $\sigma$ increases then the critical temperature of the system decreases and first order phase transitions disappear (see the right panel in Fig. 6a). Moreover, rate of decrease of the saturation value of magnetization curves increases as $h_{0}/J$ increases. On the other hand, in Fig. 6b, magnetization versus temperature curves have been plotted with $\sigma=0.5$ (left panel) and $\sigma=2.5$ (right panel) with some selected values of $h_{0}/J$. In this figure, it is clear that saturation values of magnetization curves remain unchanged for $\sigma=0.5$ and tend to decrease rapidly to zero with increasing values of $h_{0}/J$ when $\sigma=2.5$. In addition, as $h_{0}/J$ increases when the value of $\sigma$ is fixed then the critical temperature of the system decreases and ferromagnetic phase region of the system gets narrower. These observations show that there is a collective effect originating from the presence of both $h_{0}/J$ and $\sigma$ parameters on the phase diagrams and magnetization curves of the system.
\begin{figure}
\includegraphics[width=8cm]{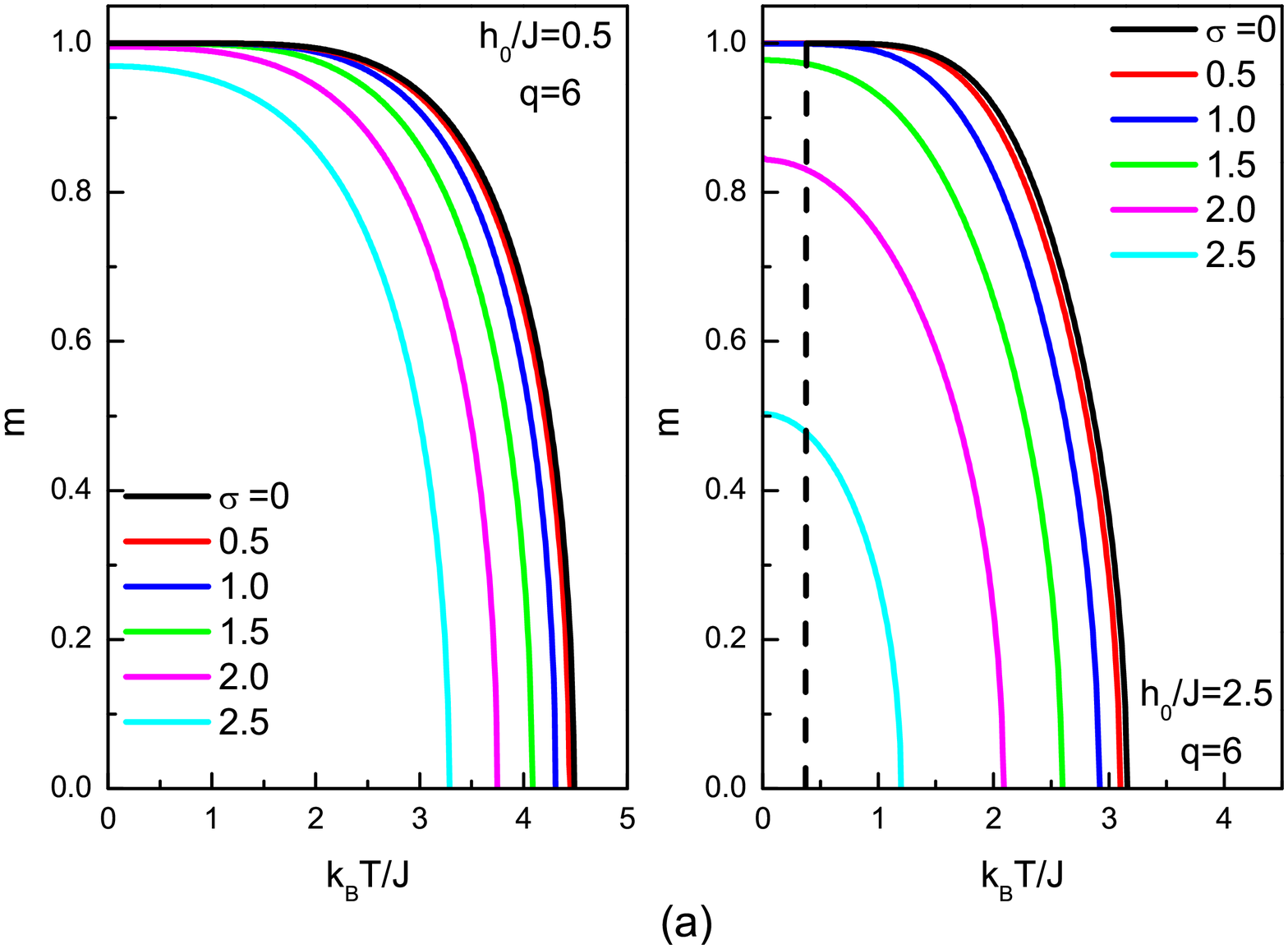}
\includegraphics[width=8cm]{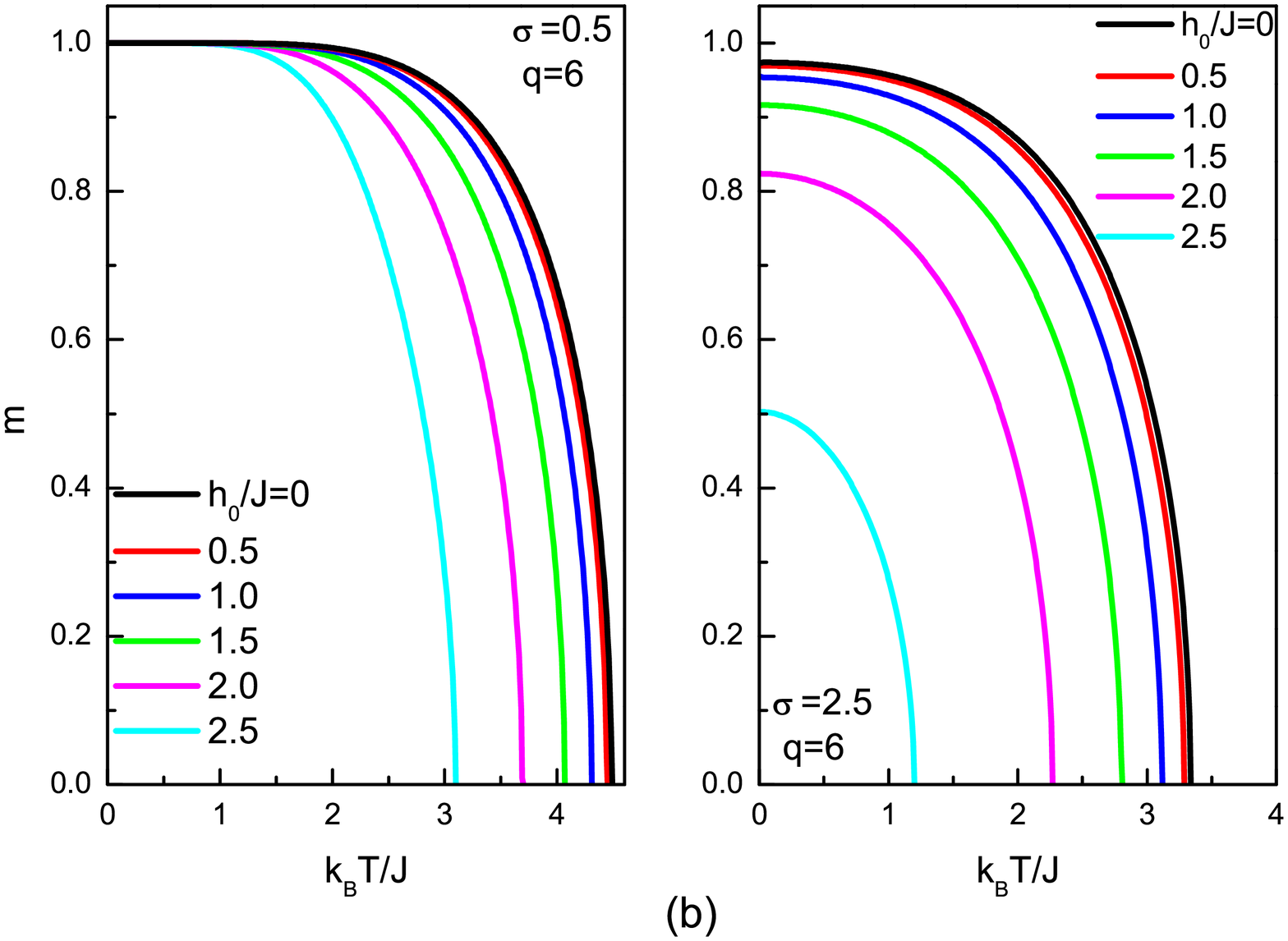}\\
\caption{Magnetization curves for simple cubic lattice corresponding to the phase diagrams in Fig. 5 for a double Gaussian distribution with some selected values of $h_{0}/J$ and $\sigma$.}\label{fig6}
\end{figure}

Finally, Fig. 7 represents the variation of the tricritical point $(k_{B}T_{t}/J, h_{t}/J)$ with $\sigma$ for $q=6, 8$ and $12$. As seen from Fig. 7, $k_{B}T_{t}/J$ value decreases monotonically as $\sigma$ increases and reaches to zero at a critical value $\sigma_{t}$. According to our calculations, the critical distribution width $\sigma_{t}$ value can be given as $\sigma_{t}=1.421, 2.238, 3.985$ for $q=6, 8$ and $12$, respectively. This implies that $\sigma_{t}$ value depends on the coordination number of the lattice. Besides, $h_{t}/J$ curves exhibit relatively small increment with increasing $\sigma$ value.
\begin{figure}
\includegraphics[width=15cm]{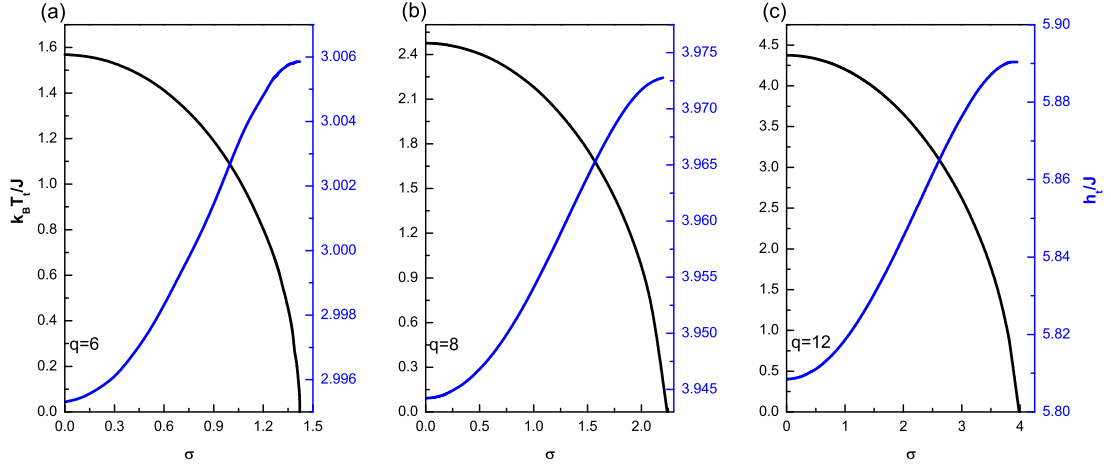}\\
\caption{Variations of tricritical temperature $k_{B}T_{t}/J$ and tricritical field $h_{t}/J$ as function of distribution width $(\sigma)$ for (a) simple cubic, (b) body centered cubic and (c) face centered cubic lattices. }\label{fig7}
\end{figure}

\section{Conclusions}\label{conclude}
In this work, we have studied the phase diagrams of a spin-$1/2$ Ising model in a random magnetic field on simple cubic, body centered cubic and face centered cubic lattices. RFIM with nearest neighbor interactions with double Gaussian distribution as well as its single Gaussian counterpart has not been examined within the EFT approximation before. Hence, we have introduced an effective field approximation that takes into account the correlations between different spins in the cluster of considered lattice and examined the phase diagrams as well as magnetization curves of the system for different types of random field distributions, namely single Gaussian, bimodal and double Gaussian distributions. For a single Gaussian distribution, we have found that the system always undergoes a second order phase transition between paramagnetic and ferromagnetic phases. In the absence of any randomness (i.e. $h_{0}/J=0, \sigma=0$) critical temperature values corresponding to the coordination numbers $q=6, 8$ and $12$ are found to be in an excellent agreement with those of the other theoretical methods such as MC and SE. For bimodal and double Gaussian distributions, we have given the proper phase diagrams, especially the first-order transition lines that include reentrant phase transition regions. Our numerical analysis clearly indicate that such field distributions lead to a tricritical behavior. Moreover, we have discussed a collective effect which arise from the presence of both $h_{0}/J$ and $\sigma$ parameters together and we have observed that saturation values of magnetization curves are strongly related to this collective effect.

As a result, we can conclude that all of the points mentioned above show that our method is superior to the conventional EFT methods based on decoupling approximation
and therefore we think that the calculation results are more accurate. Therefore, we hope that the results obtained in this work may be beneficial from both theoretical and experimental points of view.

\section*{Acknowledgements}
One of the authors (YY) would like to thank the Scientific and
Technological Research Council of Turkey (T\"{U}B\.{I}TAK) for
partial financial support. This work has been completed at the Dokuz
Eyl\"{u}l University, Graduate School of Natural and Applied
Sciences and is the subject of the forthcoming Ph.D. thesis of Y.
Y\"{u}ksel. The partial financial support from SRF
(Scientific Research Fund) of Dokuz Eyl\"{u}l University
(2009.KB.FEN.077) is also acknowledged.

\section*{Appendix}
The number of distinct correlations for the spin-$1/2$ system with $q$ nearest neighbors is $2q-1$. With the magnetization
of the central spin $\langle\langle S_{0}\rangle\rangle_{r}$ and its nearest neighbor spin $\langle\langle S_{1}\rangle\rangle_{r}$, there has to be $2q+1$ unknowns which forms a system of linear equations. $q$ of the correlations which include central spin are derived from the central spin magnetization $\langle\langle S_{0}\rangle\rangle_{r}$, and rest of the $q-1$ correlations which include only perimeter spin are derived from perimeter spin magnetization expression $\langle\langle S_{1}\rangle\rangle_{r}$. Let us label these correlations with $x_i, i=1,2,\ldots,2q+1$, such that first $q$ of them include only perimeter spins and last $q+1$ of them include central spin. We can represent all of the spin correlations and centeral/perimeter spin magnetization with $x_i$ as $x_{1}=\langle\langle S_{1}\rangle\rangle_{r}$, $x_2=\langle\langle S_{1}S_{2}\rangle\rangle_{r}$,\ldots $x_q=\langle\langle S_{1}S_{2}\ldots S_{q}\rangle\rangle_{r}$, $x_{q+1}=\langle\langle S_{0}\rangle\rangle_{r}$, $x_{q+2}=\langle\langle S_{0}S_{1}\rangle\rangle_{r},\ldots$ and $x_{2q+1}=\langle\langle S_{0}S_{1}\ldots S_{q}\rangle\rangle_{r}$. The complete set of correlation functions for $q=8,12$ can be obtained as the same procedure given between equations (\ref{eq4})-(\ref{eq11}) in section \ref{formulation}.

The complete set of correlation functions for body centered cubic lattice $(q=8)$:
\begin{eqnarray}
\nonumber
x_1&=&a_1+a_2x_9,\\
\nonumber
x_2&=&a_1x_1+a_2x_{10},\\
\nonumber
x_3&=&a_1x_2+a_2x_{11},\\
\nonumber
x_4&=&a_1x_3+a_2x_{12},\\
\nonumber
x_5&=&a_1x_4+a_2x_{13},\\
\nonumber
x_6&=&a_1x_5+a_2x_{14},\\
\nonumber
x_7&=&a_1x_6+a_2x_{15},\\
\nonumber
x_8&=&a_1x_7+a_2x_{16},\\
\nonumber
x_9&=&k_0+8k_1x_1+28k_2x_2+56k_3x_3+70k_4x_4+56k_5x_5+28k_6x_6+8k_7x_7+k_8x_8,\\
\nonumber
x_{10}&=&8k_1+(28k_2+k_0)x_1+56k_3x_2+70k_4x_3+56k_5x_4+28k_6x_5+8k_7x_6+k_8x_7,\\
\nonumber
x_{11}&=&(56k_3+8k_1)x_1+(70k_4+28k_2+k_0)x_2+56k_5x_3+28k_6x_4+8k_7x_5+k_8x_6,\\
\nonumber
x_{12}&=&(56k_3+56k_5+8k_1)x_2+(70k_4+28k_2+k_0+28k_6)x_3+8k_7x_4+k_8x_5,\\
\nonumber
x_{13}&=&(56k_3+56k_5+8k_7+8k_1)x_3+(70k_4+k_8+28k_2+k_0+28k_6)x_4,\\
\nonumber
x_{14}&=&(56k_3+56k_5+8k_7+8k_1)x_4+(70k_4+k_8+28k_2+k_0+28k_6)x_5,\\
\nonumber
x_{15}&=&(56k_3+56k_5+8k_7+8k_1)x_5+(70k_4+k_8+28k_2+k_0+28k_6)x_6,\\
\nonumber
x_{16}&=&(56k_3+56k_5+8k_7+8k_1)x_6+(70k_4+k_8+28k_2+k_0-28k_6)x_7,\\
\nonumber
x_{17}&=&(56k_3+56k_5+8k_7+8k_1)x_7+(70k_4+k_8+28k_2+k_0+28k_6)x_8.
\end{eqnarray}

$$$$
Similarly, correlation functions for face centered cubic lattice $(q=12)$:
\begin{eqnarray}
\nonumber
x_{1}&=&a_1+a_2x_{13},\\
\nonumber
x_{2}&=&a_1x_{1}+a_2x_{14},\\
\nonumber
x_{3}&=&a_1x_{2}+a_2x_{15},\\
\nonumber
x_{4}&=&a_1x_{3}+a_2x_{16},\\
\nonumber
x_{5}&=&a_1x_{4}+a_2x_{17},\\
\nonumber
x_{6}&=&a_1x_{5}+a_2x_{18},\\
\nonumber
x_{7}&=&a_1x_{6}+a_2x_{19},\\
\nonumber
x_{8}&=&a_1x_{7}+a_2x_{20},\\
\nonumber
x_{9}&=&a_1x_{8}+a_2x_{21},\\
\nonumber
x_{10}&=&a_1x_{9}+a_2x_{22},\\
\nonumber
x_{11}&=&a_1x_{10}+a_2x_{23},\\
\nonumber
x_{12}&=&a_1x_{11}+a_2x_{24},\\
\nonumber
x_{13}&=&k_0+12k_1x_{1}+66k_2x_{2}+220k_3x_{3}+495k_4x_{4}+792k_5x_{5}+924k_6x_{6}\\
\nonumber
&&+792k_7x_{7}+495k_8x_{8}+220k_9x_{9}+66k_{10}x_{10}+12k_{11}x_{11}+k_{12}x_{12},\\
\nonumber
x_{14}&=&12k_1+(k_0+66k_2)x_{1}+220k_3x_{2}+495k_4x_{3}+792k_5x_{4}+924k_6x_{5},\\
\nonumber
&&+792k_7x_{6}+495k_8x_{7}+220k_9x_{8}+66k_{10}x_{9}+12k_{11}x_{10}+k_{12}x_{11},\\
\nonumber
x_{15}&=&(12k_1+220k_3)x_{1}+(k_0+66k_2+495k_4)x_{2}+792k_5x_{3}+924k_6x_{4},\\
\nonumber
&&+792k_7x_{5}+495k_8x_{6}+220k_9x_{7}+66k_{10}x_{8}+12k_{11}x_{9}+k_{12}x_{10},\\
\nonumber
x_{16}&=&(12k_1+220k_3+792k_5)x_{2}+(k_0+66k_2+495k_4+924k_6)x_{3},\\
\nonumber
&&+792k_7x_{4}+495k_8x_{5}+220k_9x_{6}+66k_{10}x_{7}+12k_{11}x_{8}+k_{12}x_{9},\\
\nonumber
x_{17}&=&(12k_1+220k_3+792k_5+792k_7)x_{3}+(k_0+66k_2+495k_4+495k_8+924k_6)x_{4},\\
\nonumber
&&+220k_9x_{5}+66k_{10}x_{6}+12k_{11}x_{7}+k_{12}x_{8},\\
\nonumber
x_{18}&=&(12k_1+220k_3+220k_9+792k_5+792k_7)x_{4},\\
\nonumber
&&+(k_0+66k_2+495k_4+495k_8+66k_{10}+924k_6)x_{5}+12k_{11}x_{6}+k_{12}x_{7},\\
\nonumber
x_{19}&=&(12k_1+220k_3+220k_9+792k_5+12k_{11}+792k_7)x_{5},\\
\nonumber
&&+(k_0+66k_2+k_{12}+495k_4+495k_8+66k_{10}+924k_6)x_{6},\\
\nonumber
x_{20}&=&(12k_1+220k_3+220k_9+792k_5+12k_{11}+792k_7)x_{6},\\
\nonumber
&&+(k_0+66k_2+k_{12}+495k_4+495k_8+66k_{10}+924k_6)x_{7},\\
\nonumber
x_{21}&=&(12k_1+220k_3+220k_9+792k_5+12k_{11}+792k_7)x_{7},\\
\nonumber
&&+(k_0+66k_2+k_{12}+495k_4+495k_8+66k_{10}+924k_6)x_{8},\\
\nonumber
x_{22}&=&(12k_1+220k_3+220k_9+792k_5+12k_{11}+792k_7)x_{8},\\
\nonumber
&&+(k_0+66k_2+k_{12}+495k_4+495k_8+66k_{10}+924k_6)x_{9},\\
\nonumber
x_{23}&=&(12k_1+220k_3+220k_9+792k_5+12k_{11}+792k_7)x_{9},\\
\nonumber
&&+(k_0+66k_2+k_{12}+495k_4+495k_8+66k_{10}+924k_6)x_{10},\\
\nonumber
x_{24}&=&(12k_1+220k_3+220k_9+792k_5+12k_{11}+792k_7)x_{10},\\
\nonumber
&&+(k_0+66k_2+k_{12}+495k_4+495k_8+66k_{10}+924k_6)x_{11},\\
\nonumber
x_{25}&=&(12k_1+220k_3+220k_9+792k_5+12k_{11}+792k_7)x_{11},\\
\nonumber
&&+(k_0+66k_2+k_{12}+495k_4+495k_8+66k_{10}+924k_6)x_{12}.
\end{eqnarray}
where the coefficients are given by
\begin{eqnarray}\nonumber
\nonumber a_{0}=\cosh(J\nabla)F(x+\gamma)|_{x=0}, \ \ \ \ \ \ \ \ \ \ && a_{1}=\sinh(J\nabla)F(x+\gamma)|_{x=0},
\end{eqnarray}
with $\gamma=(q-1)A$,
\begin{equation}
k_{i}=\cosh^{q-i}(J\nabla)\sinh^{i}(J\nabla)F(x)|_{x=0}, \ \ i=0,\ldots ,q
\end{equation}
where $q$ is the coordination number. Since the first $2q$ equations are independent of the correlation labeled $x_{2q+1}$ it is not necessary to include this correlation function in the set of linear equations. Therefore it is adequate to take $2q$ linear equations for calculations.

\newpage
%\section*{References}


\begin{thebibliography}{10}

\expandafter\ifx\csname natexlab\endcsname\relax\def\natexlab#1{#1}\fi
\expandafter\ifx\csname bibnamefont\endcsname\relax
  \def\bibnamefont#1{#1}\fi
\expandafter\ifx\csname bibfnamefont\endcsname\relax
  \def\bibfnamefont#1{#1}\fi
\expandafter\ifx\csname citenamefont\endcsname\relax
  \def\citenamefont#1{#1}\fi
\expandafter\ifx\csname url\endcsname\relax
  \def\url#1{\texttt{#1}}\fi
\expandafter\ifx\csname urlprefix\endcsname\relax\def\urlprefix{URL }\fi
\providecommand{\bibinfo}[2]{#2}
\providecommand{\eprint}[2][]{\url{#2}}


%%%%%%%%%%%%%%%%%%%%%%%%%%%%%%%%%%%%%%%%%%%%%%%%%%%%%%%%%%%%%%%%%%%%%%%%%%%%%%%%%%%%%%%%%%%%%%%%%%%%%%%%%%%%%%%%%%%%%%%%%%
\bibitem{ising} E. Ising,  Z. Phys. \textbf{31}, 253 (1925).
\bibitem{onsager} L. Onsager, Phys. Rev. \textbf{65}, 117 (1944) .
%%%%%%%%%%%%%%%%%%%%%%%%%%%%%%%%%%%%%%%%%%%%%%%%%%%%%%%%%%%%%%%%%%%%%%%%%%%%%%%%%%%%%%%%%%%%%%%%%%%%%%%%%%%%%%%%%%%%%%%%%%

%%%%%%%%%%%%%%%%%%%%%%%%%%%%%%%%%%%%%%%%%%%%%%%%%%%%%%%%%%%%%%%%%%%%%%%%%%%%%%%%%%%%%%%%%%%%%%%%%%%%%%%%%%%%%%%%%%%%%%%%%%
\bibitem{larkin} A. I. Larkin, Sov. Phys. JETP \textbf{31}, 784 (1970).
%%%%%%%%%%%%%%%%%%%%%%%%%%%%%%%%%%%%%%%%%%%%%%%%%%%%%%%%%%%%%%%%%%%%%%%%%%%%%%%%%%%%%%%%%%%%%%%%%%%%%%%%%%%%%%%%%%%%%%%%%%

%%%%%%%%%%%%%%%%%%%%%%%%%%%%%%%%%%%%%%%%%%%%%%%%%%%%%%%%%%%%%%%%%%%%%%%%%%%%%%%%%%%%%%%%%%%%%%%%%%%%%%%%%%%%%%%%%%%%%%%%%%
\bibitem{imry_ma} Y. Imry and S. Ma, Phys. Rev. Lett. \textbf{35}, 1399 (1975).
%%%%%%%%%%%%%%%%%%%%%%%%%%%%%%%%%%%%%%%%%%%%%%%%%%%%%%%%%%%%%%%%%%%%%%%%%%%%%%%%%%%%%%%%%%%%%%%%%%%%%%%%%%%%%%%%%%%%%%%%%%

%dc=2
%%%%%%%%%%%%%%%%%%%%%%%%%%%%%%%%%%%%%%%%%%%%%%%%%%%%%%%%%%%%%%%%%%%%%%%%%%%%%%%%%%%%%%%%%%%%%%%%%%%%%%%%%%%%%%%%%%%%%%%%%%
\bibitem{grinstein} G. Grinstein and S. Ma, Phys. Rev. Lett. \textbf{49}, 685 (1982).
\bibitem{fernandez} J. F. Fernandez, G. Grinstein, Y. Imry, and S. Kirkpatrick, Phys. Rev. Lett. \textbf{51}, 203 (1983).
\bibitem{imbrie} J. Z. Imbrie, Phys. Rev. Lett. \textbf{53}, 1747 (1984).
\bibitem{bricmont} J. Bricmont and A. Kupiainen, Phys. Rev. Lett. \textbf{59}, 1829 (1987).
%%%%%%%%%%%%%%%%%%%%%%%%%%%%%%%%%%%%%%%%%%%%%%%%%%%%%%%%%%%%%%%%%%%%%%%%%%%%%%%%%%%%%%%%%%%%%%%%%%%%%%%%%%%%%%%%%%%%%%%%%%

%dc=3
%%%%%%%%%%%%%%%%%%%%%%%%%%%%%%%%%%%%%%%%%%%%%%%%%%%%%%%%%%%%%%%%%%%%%%%%%%%%%%%%%%%%%%%%%%%%%%%%%%%%%%%%%%%%%%%%%%%%%%%%%%
\bibitem{parisi} G. Parisi and N. Sourlas, Phys. Rev. Lett. \textbf{43}, 744 (1979).
\bibitem{binder} K. Binder, Y. Imry, and E. Pytte, Phys. Rev. B \textbf{24}, 6736 (1981).
\bibitem{mukamel2} E. Pytte, Y. Imry, and D. Mukamel, Phys. Rev. Lett. \textbf{46}, 1173 (1981).
\bibitem{mukamel} D. Mukamel and E. Pytte, Phys. Rev. B \textbf{25}, 4779 (1982).
\bibitem{niemi} A. Niemi, Phys. Rev. Lett. \textbf{49}, 1808 (1982).
%%%%%%%%%%%%%%%%%%%%%%%%%%%%%%%%%%%%%%%%%%%%%%%%%%%%%%%%%%%%%%%%%%%%%%%%%%%%%%%%%%%%%%%%%%%%%%%%%%%%%%%%%%%%%%%%%%%%%%%%%%

%d=2
%%%%%%%%%%%%%%%%%%%%%%%%%%%%%%%%%%%%%%%%%%%%%%%%%%%%%%%%%%%%%%%%%%%%%%%%%%%%%%%%%%%%%%%%%%%%%%%%%%%%%%%%%%%%%%%%%%%%%%%%%%
\bibitem{frontera} C. Frontera and E. Vives, Phys. Rev. E \textbf{59}, R1295 (1998).
%\bibitem{seppala} Sepp\"{a}l\"{a} E T and Alava M J \textit{Phys. Rev.} E \textbf{63} 066109
%%%%%%%%%%%%%%%%%%%%%%%%%%%%%%%%%%%%%%%%%%%%%%%%%%%%%%%%%%%%%%%%%%%%%%%%%%%%%%%%%%%%%%%%%%%%%%%%%%%%%%%%%%%%%%%%%%%%%%%%%%


%%%%%%%%%%%%%%%%%%%%%%%%%%%%%%%%%%%%%%%%%%%%%%%%%%%%%%%%%%%%%%%%%%%%%%%%%%%%%%%%%%%%%%%%%%%%%%%%%%%%%%%%%%%%%%%%%%%%%%%%%%
\bibitem{belanger} D. P. Belanger, A. R. King, and V. Jaccarino, Phys. Rev. B \textbf{31}, 4538 (1985).
\bibitem{king} A. R. King, V. Jaccarino, D. P. Belanger, and S. M. Rezende, Phys. Rev. B \textbf{32}, 503 (1985).
\bibitem{ferreira} J. B. Ferreira, A. R. King, V. Jaccarino, J. L. Cardy, and H. J. Guggenheim, Phys. Rev. B \textbf{28}, 5192 (1983).
\bibitem{yoshizawa} H. Yoshizawa, R. A. Cowley, G. Shirane, R. J. Birgenau, H. J. Guggenheim, and H. Ikeda, Phys. Rev. Lett. \textbf{48}, 438 (1982).
%%%%%%%%%%%%%%%%%%%%%%%%%%%%%%%%%%%%%%%%%%%%%%%%%%%%%%%%%%%%%%%%%%%%%%%%%%%%%%%%%%%%%%%%%%%%%%%%%%%%%%%%%%%%%%%%%%%%%%%%%%

%%%%%%%%%%%%%%%%%%%%%%%%%%%%%%%%%%%%%%%%%%%%%%%%%%%%%%%%%%%%%%%%%%%%%%%%%%%%%%%%%%%%%%%%%%%%%%%%%%%%%%%%%%%%%%%%%%%%%%%%%%
\bibitem{fishman} S. Fishman and A. Aharony, J. Phys. C Solid State Phys. \textbf{12}, L729 (1979).
\bibitem{cardy} J. L. Cardy, Phys. Rev. B \textbf{29}, 505 (1984).
%%%%%%%%%%%%%%%%%%%%%%%%%%%%%%%%%%%%%%%%%%%%%%%%%%%%%%%%%%%%%%%%%%%%%%%%%%%%%%%%%%%%%%%%%%%%%%%%%%%%%%%%%%%%%%%%%%%%%%%%%%

%%%%%%%%%%%%%%%%%%%%%%%%%%%%%%%%%%%%%%%%%%%%%%%%%%%%%%%%%%%%%%%%%%%%%%%%%%%%%%%%%%%%%%%%%%%%%%%%%%%%%%%%%%%%%%%%%%%%%%%%%%
\bibitem{schneider_pytte} T. Schneider and E. Pytte, Phys. Rev. B \textbf{15}, 1519 (1987).
\bibitem{aharony} A. Aharony, Phys. Rev. B \textbf{18}, 3318 (1978).
\bibitem{matthis} D. C. Matthis, Phys. Rev. Lett. \textbf{55}, 3009 (1985).
%%%%%%%%%%%%%%%%%%%%%%%%%%%%%%%%%%%%%%%%%%%%%%%%%%%%%%%%%%%%%%%%%%%%%%%%%%%%%%%%%%%%%%%%%%%%%%%%%%%%%%%%%%%%%%%%%%%%%%%%%%

%rep.
%%%%%%%%%%%%%%%%%%%%%%%%%%%%%%%%%%%%%%%%%%%%%%%%%%%%%%%%%%%%%%%%%%%%%%%%%%%%%%%%%%%%%%%%%%%%%%%%%%%%%%%%%%%%%%%%%%%%%%%%%%
\bibitem{crokidakis} N. Crokidakis and F. D. Nobre, J. Phys.: Condens. Matter \textbf{20}, 145211 (2008).
\bibitem{salmon} O. R. Salmon, N. Crokidakis, and F. D. Nobre, J. Phys.: Condens. Matter \textbf{21}, 056005 (2009).
%%%%%%%%%%%%%%%%%%%%%%%%%%%%%%%%%%%%%%%%%%%%%%%%%%%%%%%%%%%%%%%%%%%%%%%%%%%%%%%%%%%%%%%%%%%%%%%%%%%%%%%%%%%%%%%%%%%%%%%%%%

%EFT
%%%%%%%%%%%%%%%%%%%%%%%%%%%%%%%%%%%%%%%%%%%%%%%%%%%%%%%%%%%%%%%%%%%%%%%%%%%%%%%%%%%%%%%%%%%%%%%%%%%%%%%%%%%%%%%%%%%%%%%%%%
\bibitem{borges} H. E. Borges and P. R. Silva, Physica A \textbf{144}, 561 (1987).
\bibitem{liang2} Y. Q. Liang, G. Z. Wei, Q. Zhang, Z. H. Xin, and G. L. Song, J. Magn. Magn. Mater. \textbf{284}, 47 (2004).
\bibitem{sarmento} E. F. Sarmento and T. Kaneyoshi, Phys. Rev. B \textbf{39}, 9555 (1989).
\bibitem{sebastianes} R. M. Sebastianes and W. Figueiredo, Phys. Rev. B \textbf{46}, 969 (1992).
\bibitem{kaneyoshi} T. Kaneyoshi, Physica A \textbf{139}, 455 (1985).
%%%%%%%%%%%%%%%%%%%%%%%%%%%%%%%%%%%%%%%%%%%%%%%%%%%%%%%%%%%%%%%%%%%%%%%%%%%%%%%%%%%%%%%%%%%%%%%%%%%%%%%%%%%%%%%%%%%%%%%%%%


%MC
%%%%%%%%%%%%%%%%%%%%%%%%%%%%%%%%%%%%%%%%%%%%%%%%%%%%%%%%%%%%%%%%%%%%%%%%%%%%%%%%%%%%%%%%%%%%%%%%%%%%%%%%%%%%%%%%%%%%%%%%%%
\bibitem{landau} D. P. Landau, H. H. Lee, and W. Kao, J. Appl. Phys. \textbf{49}, 1356 (1978).
\bibitem{machta} J. Machta, M. E. J. Newman, and L. B. Chayes, Phys. Rev. E \textbf{62}, 8782 (2000).
\bibitem{fytas1} N. G. Fytas, A. Malakis, and K. Eftaxias, J. Stat. Mech. Theory Exp. \textbf{2008}, 03015 (2008).
\bibitem{fytas2} N. G. Fytas and A. Malakis, Eur. Phys. J. B \textbf{61}, 111 (2008).
%%%%%%%%%%%%%%%%%%%%%%%%%%%%%%%%%%%%%%%%%%%%%%%%%%%%%%%%%%%%%%%%%%%%%%%%%%%%%%%%%%%%%%%%%%%%%%%%%%%%%%%%%%%%%%%%%%%%%%%%%%


%PA
%%%%%%%%%%%%%%%%%%%%%%%%%%%%%%%%%%%%%%%%%%%%%%%%%%%%%%%%%%%%%%%%%%%%%%%%%%%%%%%%%%%%%%%%%%%%%%%%%%%%%%%%%%%%%%%%%%%%%%%%%%
\bibitem{albayrak} E. Albayrak and O. Canko, J. Magn. Magn. Mater. \textbf{270}, 333 (2004).
%%%%%%%%%%%%%%%%%%%%%%%%%%%%%%%%%%%%%%%%%%%%%%%%%%%%%%%%%%%%%%%%%%%%%%%%%%%%%%%%%%%%%%%%%%%%%%%%%%%%%%%%%%%%%%%%%%%%%%%%%%

%SE
%%%%%%%%%%%%%%%%%%%%%%%%%%%%%%%%%%%%%%%%%%%%%%%%%%%%%%%%%%%%%%%%%%%%%%%%%%%%%%%%%%%%%%%%%%%%%%%%%%%%%%%%%%%%%%%%%%%%%%%%%%
\bibitem{gofman} M. Gofman, J. Adler, A. Aharony, A. B. Harris, and M. Schwartz, Phys. Rev. B \textbf{53}, 6362 (1996).
%%%%%%%%%%%%%%%%%%%%%%%%%%%%%%%%%%%%%%%%%%%%%%%%%%%%%%%%%%%%%%%%%%%%%%%%%%%%%%%%%%%%%%%%%%%%%%%%%%%%%%%%%%%%%%%%%%%%%%%%%%


%%%%%%%%%%%%%%%%%%%%%%%%%%%%%%%%%%%%%%%%%%%%%%%%%%%%%%%%%%%%%%%%%%%%%%%%%%%%%%%%%%%%%%%%%%%%%%%%%%%%%%%%%%%%%%%%%%%%%%%%%%
\bibitem{polat} H. Polat, \"{U}. Ak{\i}nc{\i}, and \.{I}. S\"{o}kmen, Phys. Stat. Sol. B \textbf{240}, 189 (2003).
\bibitem{canpolat} Y. Canpolat, A. Torg\"{u}rs\"{u}l, and H. Polat, Phys. Scr. \textbf{76}, 597 (2007).
\bibitem{yuksel_1} Y. Y\"{u}ksel, \"{U}. Ak{\i}nc{\i}, and H. Polat, Phys. Scr. \textbf{79}, 045009 (2009).
\bibitem{yuksel_2} Y. Y\"{u}ksel and H. Polat, J. Magn. Magn. Mater. \textbf{322}, 3907 (2010).
\bibitem{yuksel_3} \"{U}. Ak{\i}nc{\i}, Y. Y\"{u}ksel, and H. Polat, Physica A \textbf{390}, 541 (2011).
%%%%%%%%%%%%%%%%%%%%%%%%%%%%%%%%%%%%%%%%%%%%%%%%%%%%%%%%%%%%%%%%%%%%%%%%%%%%%%%%%%%%%%%%%%%%%%%%%%%%%%%%%%%%%%%%%%%%%%%%%%

%MFT
%%%%%%%%%%%%%%%%%%%%%%%%%%%%%%%%%%%%%%%%%%%%%%%%%%%%%%%%%%%%%%%%%%%%%%%%%%%%%%%%%%%%%%%%%%%%%%%%%%%%%%%%%%%%%%%%%%%%%%%%%%
\bibitem{hadjiagapiou} I. A. Hadjiagapiou, Physica A \textbf{389}, 3945 (2010).
%%%%%%%%%%%%%%%%%%%%%%%%%%%%%%%%%%%%%%%%%%%%%%%%%%%%%%%%%%%%%%%%%%%%%%%%%%%%%%%%%%%%%%%%%%%%%%%%%%%%%%%%%%%%%%%%%%%%%%%%%%

%%%%%%%%%%%%%%%%%%%%%%%%%%%%%%%%%%%%%%%%%%%%%%%%%%%%%%%%%%%%%%%%%%%%%%%%%%%%%%%%%%%%%%%%%%%%%%%%%%%%%%%%%%%%%%%%%%%%%%%%%%
\bibitem{du} A. Du, H. J. Liu, and Y. Q. Y\"{u}, Phys. Stat. Sol. B \textbf{241}, 175 (2004).
\bibitem{kaneyoshi_2} T. Kaneyoshi, I. P. Fittipaldi, R. Honmura, and T. Manabe, Phys. Rev. B  \textbf{24}, 481 (1981).
\bibitem{balcerzak} T. Balcerzak, Physica A  \textbf{317}, 213 (2003).
\bibitem{kaneyoshi_3} T. Kaneyoshi, Rev. Solid State Sci.  \textbf{2}, 39 (1988).
\bibitem{kikuchi} R. Kikuchi, Phys. Rev. \textbf{81}, 988 (1951).
\bibitem{sousa} J. R. de Sousa and I. G. Ara\'{u}jo, J. Magn. Magn. Mater. \textbf{202}, 231 (1999);  M. A. Neto and J. R. de Sousa, Phys. Rev. E \textbf{70}, 224436 (2004).
\bibitem{reinerhr} E. E. Reinerhr and W. Figueiredo, Phys. Lett. A \textbf{244}, 165 (1998).
\bibitem{landau2} D. P. Landau, Phys. Rev. B \textbf{16}, 4164 (1977); \textbf{14},  255 (1976); A. M. Ferrenberg and D. P. Landau Phys. Rev. B \textbf{44}, 5081 (1991).
\bibitem{fisher} M. E. Fisher, Rep. Prog. Phys. \textbf{30}, 615 (1967).
%%%%%%%%%%%%%%%%%%%%%%%%%%%%%%%%%%%%%%%%%%%%%%%%%%%%%%%%%%%%%%%%%%%%%%%%%%%%%%%%%%%%%%%%%%%%%%%%%%%%%%%%%%%%%%%%%%%%%%%%%%


%%%%%%%%%%%%%%%%%%%%%%%%%%%%%%%%%%%%%%%%%%%%%%%%%%%%%%%%%%%%%%%%%%%%%%%%%%%%%%%%%%%%%%%%%%%%%%%%%%%%%%%%%%%%%%%%%%%%%%%%%%
\bibitem{callen} H. B. Callen, Phys. Lett. \textbf{4}, I61 (1963).
\bibitem{honmura_kaneyoshi} R. Honmura and T. Kaneyoshi, J. Phys. C Solid State Phys. \textbf{12}, 3979 (1979).
\bibitem{kaneyoshi_1} T. Kaneyoshi, Acta Phys. Pol. A \textbf{83}, 703 (1993).
\bibitem{tamura_kaneyoshi} I. Tamura and T. Kaneyoshi, Prog. Theor. Phys. \textbf{66}, 1892 (1981).
%%%%%%%%%%%%%%%%%%%%%%%%%%%%%%%%%%%%%%%%%%%%%%%%%%%%%%%%%%%%%%%%%%%%%%%%%%%%%%%%%%%%%%%%%%%%%%%%%%%%%%%%%%%%%%%%%%%%%%%%%%


\end{thebibliography}
\end{document}